\documentstyle[aps,prl,epsfig]{revtex}
\draft
\begin{document}

\title{Self-organized and driven phase synchronization in coupled map
scale free networks} 
\author {Sarika Jalan\footnote{e-mail: sarika@prl.ernet.in} and
R. E. Amritkar\footnote{e-mail: amritkar@prl.ernet.in}}
\address{Physical Research Laboratory, Navrangpura, Ahmedabad 380 009, India.}
\maketitle
  
\begin{abstract}
As a model of evolving networks, we study coupled logistic maps on scale
free networks. For small coupling strengths nodes show turbulent behavior
but form phase synchronized clusters as coupling increases. We identify two
different ways of cluster formation. For small coupling strengths we get
{\it self-organized clusters} which have mostly intra-cluster couplings and
for large coupling strengths there is a crossover and reorganization to
{\it driven clusters} which have mostly inter-cluster couplings. In the
novel driven synchronization the nodes of one cluster are driven by those
of the others.
\end{abstract} 
 
Recently, there is considerable interest in complex systems described by
networks or graphs with complex topology \cite{Strogatz}.
One significant recent discovery in the field of complex networks is
the observation that a number of naturally occurring large and complex networks
are scale free, i.e. the  probability that a node 
in the network is connected to $k$ other nodes decays as a power
law. These networks are found in many diverse systems such as
the nervous systems \cite{koch},
social groups  \cite{social},
world wide web \cite{www}, metabolic networks \cite{metabolic},
food webs \cite{food} and citation network \cite{citation}.
Barabasi and Albert \cite{scalefree} have given a simple growth model
based on preferential attachments for the scale free networks.

Most networks in the real world consist of dynamical
elements interacting with each other. Thus in order to understand properties of such
networks, we study a coupled map model of scale free networks.
Over the past decade, much work on coupled maps has used
regular coupling schemes. They show rich phenomenology arising when
opposing tendencies compete; the nonlinear dynamics of the maps
which in the chaotic regime tends to separate the orbits of different elements,
and the couplings that tend to synchronize them.
Coupled map lattices with nearest neighbor
or short range interactions show interesting spatio-temporal patterns,
and intermittent behaviour \cite{CML}. Globally coupled
maps (GCM) where each node is connected with all other nodes, show
interesting synchronized behaviour \cite{GCM}.
Random networks with large number of connections also show
synchronized behaviour for large coupling strengths.
\cite{random-net}.

In this letter, we study the dynamics of coupled maps on scale free
networks. In particular, we investigate the clustering and synchronization
properties of such dynamically evolving networks. We find that as the
network evolves, it splits into
several phase synchronized clusters. Phase synchronization is obviously
because of the couplings between the nodes of the network and may be
achieved in two different ways. (i) The nodes of a cluster can be
synchronized because of intra-cluster couplings. We refer to 
this as the {\it self-organized synchronization}. 
(ii) Alternately, the nodes of
a cluster can be synchronized because of inter-cluster couplings. We
refer to this as the {\it driven
synchronization}. We find examples of both these types of phase
synchronized clusters in scale free networks with a crossover and
reorganization of nodes between
the two types as the coupling strength is varied.  For small couplings
synchronization is of the self-organized type while for large
couplings it is of the driven type. We will discuss several examples
of such clusters in natural systems afterwards.

Consider a network of $N$ nodes that are coupled with each
other through connections of the scale free type. Let each node of the
network be assigned a dynamical variable $x^i, i=1,2,\ldots,N$. The
evolution of the dynamical variables is given by
\begin{equation}
x^{i}_{t + 1} = (1 - \epsilon) f( x^{i}_t ) + \frac{\epsilon}{\sum_j C_{i j}}
\sum C_{ij} g( x^{j}_t )
\label{coupleddyn}
\end{equation}
where $x^{i}_n$ is the dynamical variable of the i-th node $(1 \le i \le N)$ at
the $t-th$ time step, C is the coupling matrix with elements $C_{ij}$
taking values $1$ or $0$ depending upon whether i and j are 
connected or not.  Note that C is a symmetric matrix with diagonal 
elements zero. The function $f(x)$ defines the local nonlinear map and 
the function $g(x)$ defines the nature of coupling between the nodes. 
In this paper, we define the local dynamics by the logistic map,
$f(x) = \mu x (1 - x)$. The coupling function is taken as the 
identity mapping, $g(x)=x$.

The scale free network of $N$ nodes is generated by using the model of 
Barabasi et.al. \cite{model}.
Starting with a small number, $m_0$, of nodes, at every time 
step a new node with $m \le m_0$ connections is added. The probability
$\pi(k_i)$ that a
connection starting from this new node is connected
to a node $i$ depends on the degree
$k_i$ of node $i$ (preferential attachment) and is given by
	   $\pi(k_i) = (k_i + 1)/(\sum_j (k_j + 1))$.
After $\tau$ time steps the
model leads to a random network with $N = \tau + m_0$ nodes and $m \tau$
connections. This model leads to a scale free network, i.e. the
probability $P(k)$ that a node has a degree $k$ decays as a power law, 
$P(k) \sim k^\lambda$ where $\lambda$ is a constant.
For the type of
probability law $\pi(k)$ that we have used $\lambda=3$. Other forms
for the probability $\pi(k)$ are possible which give different
values of $\lambda$. However, the results reported in this letter do
not depend on the exact form of $\pi(k)$ except that it should lead to
a scale free network.

Synchronization of coupled dynamical systems may be defined in various
ways. The perfect synchronization corresponds to
the dynamical variables for different nodes having identical
values. The phase synchronization
corresponds to the dynamical variables for different nodes having
values with some definite relations \cite{phase}. In scale free
networks, we find that when the local dynamics of the
nodes (i.e. function $f(x)$) is in the chaotic zone, perfect
synchronization is obtained only for
clusters with small number of nodes. However,
interesting results are obtained when we study phase synchronized
behaviour. For our study we define the phase synchronization
as follows \cite{syn}.
Let $n_i$ and $n_j$ denote
the number of times the dynamical variables $x^i_t$ and $x^j_t$,
$t=1,2,\ldots,T$ for the nodes $i$ and $j$ show local 
minima during the time interval $T$. Let $n_{ij}$
denote the number of times these local minima match
with each other. We define the phase distance
between the nodes $i$ and $j$ as $d_{ij}=1-2n_{ij}/(n_i+n_j)$. Clearly, $d_{ij}=0$
when all minima of variables $x^i$ and $x^j$ match with each other
and $d_{ij}=1$ when none of the minima match.
We say that nodes $i$ and $j$ are phase synchronized if $d_{ij}=0$. 
Also, a cluster of nodes is phase synchronized if all pairs of nodes
belonging to that cluster are phase synchronized.

\begin{figure}

\begin{center}
\epsfxsize 8cm \epsfysize 8cm
\epsfbox[150 375 495 718]{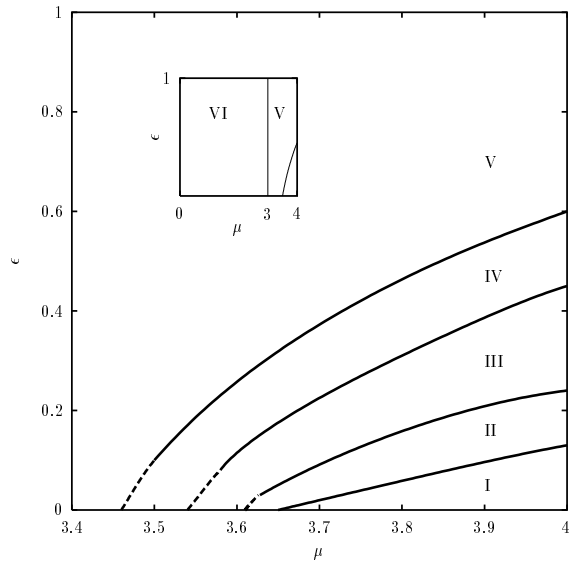}
\caption{Phase diagram showing turbulent, phase synchronized and
coherent regions in the two parameter space of $\mu$ and
$\epsilon$. Different regions are I. Turbulent region, II. Mixed
region, III. Partially ordered region, IV. Ordered quasiperiodic
region, V. Ordered periodic region, VI. Coherent region. Calculations
are for $N=50, m=1, T=100$. Region boundaries are determined based on changes
in the behaviour of the largest Lyapunov exponent (see Fig.~(2)) and
observing the asymptotic behaviour using several
initial conditions. The dashed lines indicate uncertainties in
determining the boundaries. The inset shows the phase diagram for the entire
range of parameter $\mu$ i.e. from 0 to 4.}
\end{center}
\label{phase}

\end{figure}

We now present numerical results of our model. We generate scale free
networks using the algorithm defined above and then study coupled
dynamics of variables associated with nodes of the network. Starting
from random initial conditions the dynamics of Eq.~(\ref{coupleddyn}),
after an initial transient, leads to interesting phase synchronized
behaviour. Fig.~(1) shows the phase diagram in the two parameter space
defined by $\mu$ and $\epsilon$ for $m=1, N=50, T=100$. To understand the
phase diagram let us first consider different states which are
obtained by the coupled dynamics. \\
(a) Coherent state: Nodes form a single synchronized cluster. \\
(b) Ordered state: Nodes form two or more clusters. \\
(c) Partially ordered states: Nodes form a few clusters with some
nodes not attached to any clusters. \\
(d) Turbulent state: All nodes behave chaotically with no cluster
formation. \\

\begin{figure}
\begin{center}
\epsfxsize 8cm \epsfysize 8cm
\epsfbox[146 373 490 712]{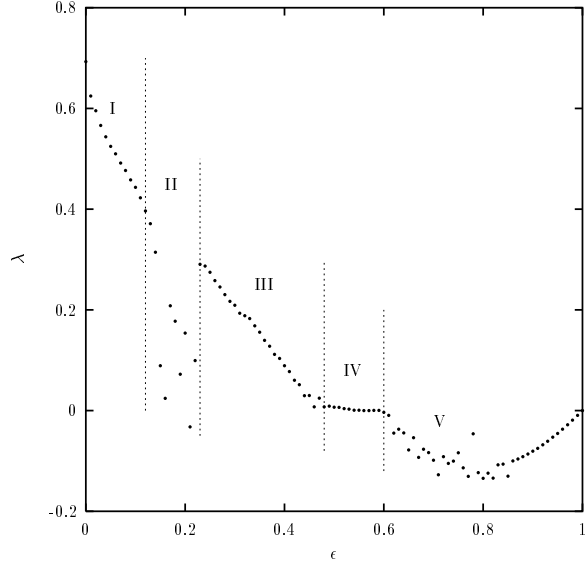}
\caption{Largest Lyapunov exponent, $\lambda$, is plotted as a function of
$\epsilon$ for $\mu = 4.0$. Different regions are labeled as in Fig.~(1).}
\label{lya}
\end{center}
\end{figure}

For $\mu<3$, we get a stable coherent region (region VI) with all nodes having the
fixed point value. To understand the remaining phase diagram, consider
the line $\mu=4$. Fig.~(2) shows the largest Lyapunov exponent
$\lambda$ as a function of the coupling strength $\epsilon$ for
$\mu=4$. We can identify five different regions
as $\epsilon$ increases from 0 to 1; namely turbulent region, mixed
region, partially ordered region, ordered quasiperiodic region and
ordered periodic region as shown by regions I to V in Figs.~(1) and~(2).
For small values of $\epsilon$, we observe a turbulent
behaviour with all nodes evolving chaotically and there is no phase
synchronization. As $\epsilon$ increases further we get into a mixed
region (region II) which shows variety of phase synchronized behaviour, namely
ordered chaotic, ordered quasiperiodic,
ordered periodic and partially ordered, depending on the initial
conditions. Next region (region III) shows partially ordered chaotic behaviour. Here, 
the number of clusters as well as the number of nodes in the clusters depend on the
initial conditions and also they change with time. There are
several isolated nodes not belonging to any cluster. Many of these
nodes are of the floating type which keep on switching intermittently
between independent evolution and phase synchronized evolution
attached to some cluster. 
Last two regions are ordered quasiperiodic
and ordered periodic regions. In these regions, the network always splits into 
two clusters. The two clusters are perfectly anti-phase synchronized
with each other, i.e. when the nodes belonging to one cluster show
minima those belonging to the other cluster show maxima.

We now investigate the nature of phase ordering in different regions
of the phase diagram. We first concentrate on the middle of 
regions II and V where we observe ordered
periodic behaviour. In both cases the largest Lyapunov
exponent is negative. In Fig.~(3a) and~(3b) we show the coupling matrix
$C$ (solid circles) and the nodes belonging to the two clusters
in regions II and V respectively. In Fig.~(3a) we observe that
there are no inter-cluster couplings between the nodes of the two clusters except
one coupling i.e. all the couplings except one are of the
intra-cluster type. The phase synchronization in this case is clearly of the
{\it self-organized} type. Exactly opposite behaviour is observed for the
region V (Fig.~(3b)). Here, we find that all the couplings are of inter-cluster 
type with no intra-cluster couplings. This is clearly the phenomena of {\it driven 
synchronization} where nodes of one cluster are driven into a phase
synchronized state due to the couplings with nodes of the other
cluster. The phenomena of driven synchronization in this region is a very robust one
in the sense that it is obtained for almost all initial conditions, the
transient time is very small, the nodes belonging to the two clusters
are uniquely determined and we get a stable solution.

We observe that for small values of $\epsilon$ the self organized
behaviour dominates while for large $\epsilon$ driven behaviour dominates.
As the coupling parameter $\epsilon$ increases from zero and we enter
region II, we observe phase synchronized clusters of the self organized
type. Region III acts as a
crossover region from the self-organized to the driven behaviour. Here, the
clusters are of mixed type. The number of inter-cluster couplings
is approximately same as the number of intra-cluster couplings.
In this region there is a competition between the self-organized and driven 
behaviour. This appears to be the reason for the formation of several
clusters and floating nodes as well as the sensitivity of these to the 
initial conditions. As $\epsilon$ increases, we get into region IV where the driven
synchronization dominates and most of the connections between the nodes
are of the inter-cluster type. This driven
synchronization is further stabilized in region V with two
perfectly anti-phase synchronized driven clusters.

\begin{figure}
\begin{center}
\epsfxsize 8cm \epsfysize 8cm
\epsfbox[145 352 485 714]{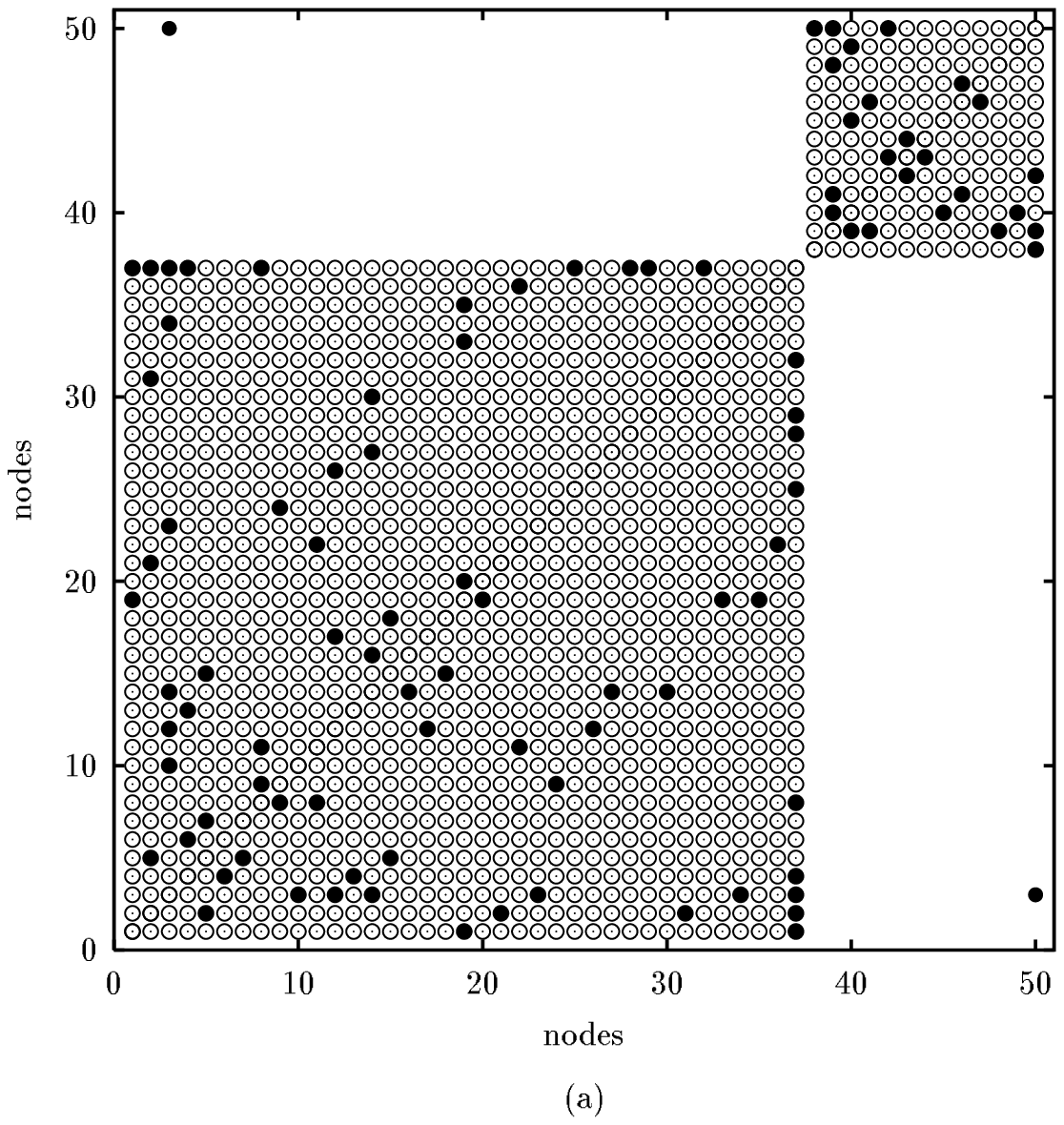}
\epsfxsize 8cm \epsfysize 8cm
\epsfbox[145 352 485 714]{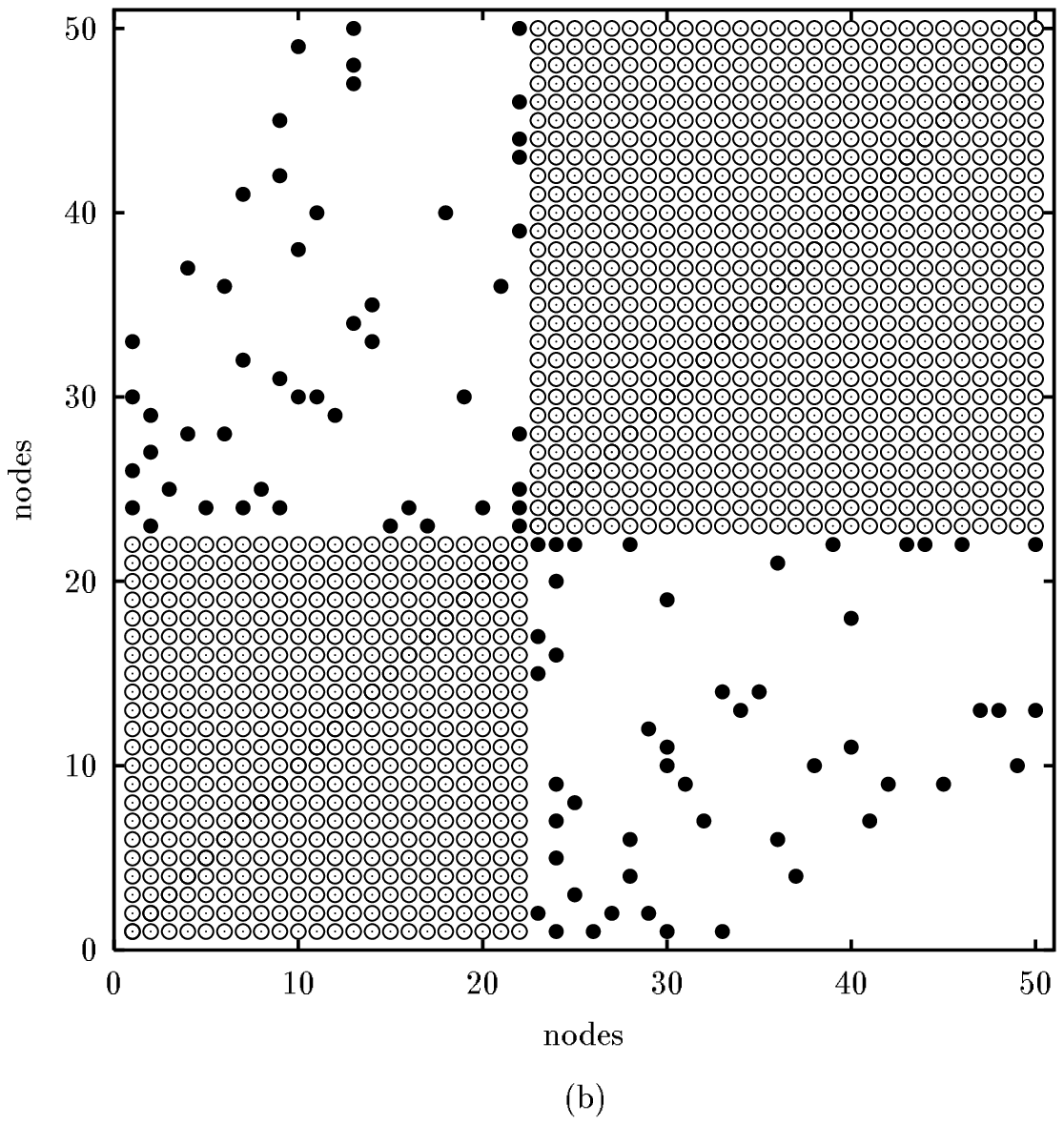}
\caption{Figures (a) and (b) show formation of phase synchronized
clusters for $\mu=4$ and $\epsilon=0.16$ and $0.8$ respectively. For
clarity of display, the
nodes are renumbered in each case such that nodes belonging to the
same cluster have sequential numbers. The nodes of clusters are shown
by open circles in squares. The elements of the coupling
matrix are shown by solid circles.}
\end{center}
\label{cluster}
\end{figure}

Geometrically, the organization of the network into connections of both
self-organized and driven types is always possible for $m=1$. For
$m=1$, our growth algorithm generates a tree type structure. A tree can 
always be broken into two or more disjoint clusters with only
intra-cluster couplings by breaking one
or more connections. Clearly, this splitting is not unique. A tree can 
also be divided into two clusters by putting connected nodes into different
clusters. This division is unique and leads to two clusters with only
inter-cluster couplings.

For $m >1$ the dynamics of Eq.~(\ref{coupleddyn}) leads to a similar
phase diagram as in Fig~(1) with region II dominated by
self-organized synchronization and regions IV and V dominated by driven
synchronization. Though perfect inter- and
intra-cluster couplings between 
the nodes as displayed in Figs.~(3a) and~(3b) are no longer
observed, clustering in region II is such that most of the couplings are
of intra-cluster type while for regions IV and V they 
are of the inter-cluster type. As $m$ increases the
regions I and II are mostly
unaffected, but regions IV and V shrink while region III grows in size.

The phenomena of self-organized and driven behaviour persists for the
largest size network that we have studied ($N=1000$). The mixed region 
showing self-organized behaviour is mostly unaffected while the
ordered regions showing driven behaviour show a small shrinking in size.

We have generated the networks using the preferential attachment law
for $\pi(k_i)$ leading to a power law distribution of the
degree of nodes. If instead we use a non-preferential law of
attachment i.e. the probability of connecting to any node does not 
depend on the degree of that node but is a constant, then the
distribution of degree of nodes shows
an exponential dependence rather than a power law. In this case we
do not observe any phase synchronized behaviour. Thus the scale free
nature appears to be important for observing the phase synchronized
behaviour.

There are several examples of self-organized and driven behaviour in
naturally occurring systems. Self-organized behaviour is more common
and is easily observed. Examples are social, ethnic and religious
groups, political groups, cartel of
industries and countries, herds of animals and flocks of birds,
different dynamic transitions such as self-organized criticality etc. The
driven behaviour is not so common. An interesting example is the
behaviour of fans during a match between traditional rivals. 
Before the match the fans may act as individuals
(turbulent behavior) or form self-organized clusters such as a
single cluster of fans of the game or several clusters of fans of
different star players. During the match there can be
a crossover to a driven behaviour. When the
match reaches a feverish pitch, i.e. the strength of the interaction
increases, the fans are likely to form two phase synchronized groups.
The response of the two groups will be
anti-phase synchronized with each other. Another example of crossover
to a driven behaviour is the conflict in Bosnia where a society organized
into villages and towns was split into ethnic groups.

As discussed in the introduction several naturally occurring networks
show scale free behaviour and it is likely to be a generic behaviour
of several naturally growing networks. We expect to observe both the self-organized
and driven synchronization behaviour reported in this paper in such
systems. Network properties of many examples of self-organized and
driven behaviour discussed above
are not known and it is possible that some of these
examples may have scale free type of networks.

To conclude we have found interesting self-organized and driven phase
synchronization behaviour in coupled maps on scale free
networks. Self-organized synchronization is
characterized by dominant intra-cluster couplings and is found when
strength of the couplings is small as compared to the local dynamics.
As the coupling strength increases there is a crossover from
the self-organized to the driven behaviour which also involves
reorganization of nodes into different clusters.
The driven behaviour is characterized by inter-cluster couplings
and is found when strength of the couplings is large as compared to
the local dynamics.

One of us (REA) thanks Professor H. Kanz for useful
discussions and Max-Plank Institute for
the Physics of Complex Systems,  Dresden for hospitality.

\end{document}